\documentclass[sigconf]{acmart}
\AtBeginDocument{%
  \providecommand\BibTeX{{%
    \normalfont B\kern-0.5em{\scshape i\kern-0.25em b}\kern-0.8em\TeX}}}

\copyrightyear{2024}
\acmYear{2024}
\setcopyright{rightsretained}
\acmConference[CHI EA '24]{Extended Abstracts of the CHI Conference on Human Factors in Computing Systems}{May 11--16, 2024}{Honolulu, HI, USA}
\acmBooktitle{Extended Abstracts of the CHI Conference on Human Factors in Computing Systems (CHI EA '24), May 11--16, 2024, Honolulu, HI, USA}
\acmDOI{10.1145/3613905.3650740}
\acmISBN{979-8-4007-0331-7/24/05}

\usepackage{caption}
\usepackage{subcaption}
\usepackage{longtable}
\usepackage{soul}
\begin{document}


\title[Assistive Technology for CVI]{Broadening Our View: Assistive Technology for Cerebral Visual Impairment}



\author{Bhanuka Gamage}
\email{bhanuka.gamage@monash.edu}
\orcid{0000-0003-0502-5883}
\affiliation{%
  \institution{Monash University}
  \streetaddress{Wellington Rd}
  \city{Melbourne}
  \country{Australia}
  \postcode{3800}
}

\author{Leona Holloway}
\email{leona.holloway@monash.edu}
\orcid{0000-0001-9200-5164}
\affiliation{%
  \institution{Monash University}
  \streetaddress{Wellington Rd}
  \city{Melbourne}
  \country{Australia}
  \postcode{3800}
}

\author{Nicola McDowell}
\email{n.mcdowell@massey.ac.nz}
\orcid{0000-0001-6969-9604}
\affiliation{%
  \institution{Massey University}
  \city{Auckland}
  \country{New Zealand}
}

\author{Thanh-Toan Do}
\email{toan.do@monash.edu}
\orcid{0000-0002-6249-0848}
\affiliation{%
  \institution{Monash University}
  \streetaddress{Wellington Rd}
  \city{Melbourne}
  \country{Australia}
  \postcode{3800}
}

\author{Nicholas Price}
\email{nicholas.price@monash.edu}
\orcid{0000-0001-9404-7704}
\affiliation{%
  \institution{Monash University}
  \streetaddress{Wellington Rd}
  \city{Melbourne}
  \country{Australia}
  \postcode{3800}
}

\author{Arthur Lowery}
\email{arthur.lowery@monash.edu}
\orcid{0000-0001-7237-0121}
\affiliation{%
  \institution{Monash University}
  \streetaddress{Wellington Rd}
  \city{Melbourne}
  \country{Australia}
  \postcode{3800}
}

\author{Kim Marriott}
\orcid{0000-0002-9813-0377}
\email{kim.marriott@monash.edu}
\affiliation{%
  \institution{Monash University}
  \streetaddress{Wellington Rd}
  \city{Melbourne}
  \country{Australia}
  \postcode{3800}
}

\renewcommand{\shortauthors}{Bhanuka Gamage, et al.}

\begin{abstract}
Over the past decade, considerable research has been directed towards assistive technologies to support people with vision impairments using machine learning, computer vision, image enhancement, and/or augmented/virtual reality.
However, this has almost totally overlooked a growing demographic: people with Cerebral Visual Impairment (CVI). 
Unlike Ocular Vision Impairments (OVI), CVI arises from damage to the brain's visual processing centres.
This paper introduces CVI and reveals a wide research gap in addressing the needs of this demographic. 
Through a scoping review, we identified 14 papers at the intersection of these technologies and CVI. 
Of these, only three papers described assistive technologies focused on people living with CVI, with the others focusing on diagnosis, understanding, simulation or rehabilitation. 
Our findings highlight the opportunity for the Human-Computer Interaction and Assistive Technologies research community to explore and address this underrepresented domain, thereby enhancing the quality of life for people with CVI.
\end{abstract}





\maketitle

\begin{figure*}[!t]
    \centering
    \includegraphics[width=14cm,keepaspectratio]{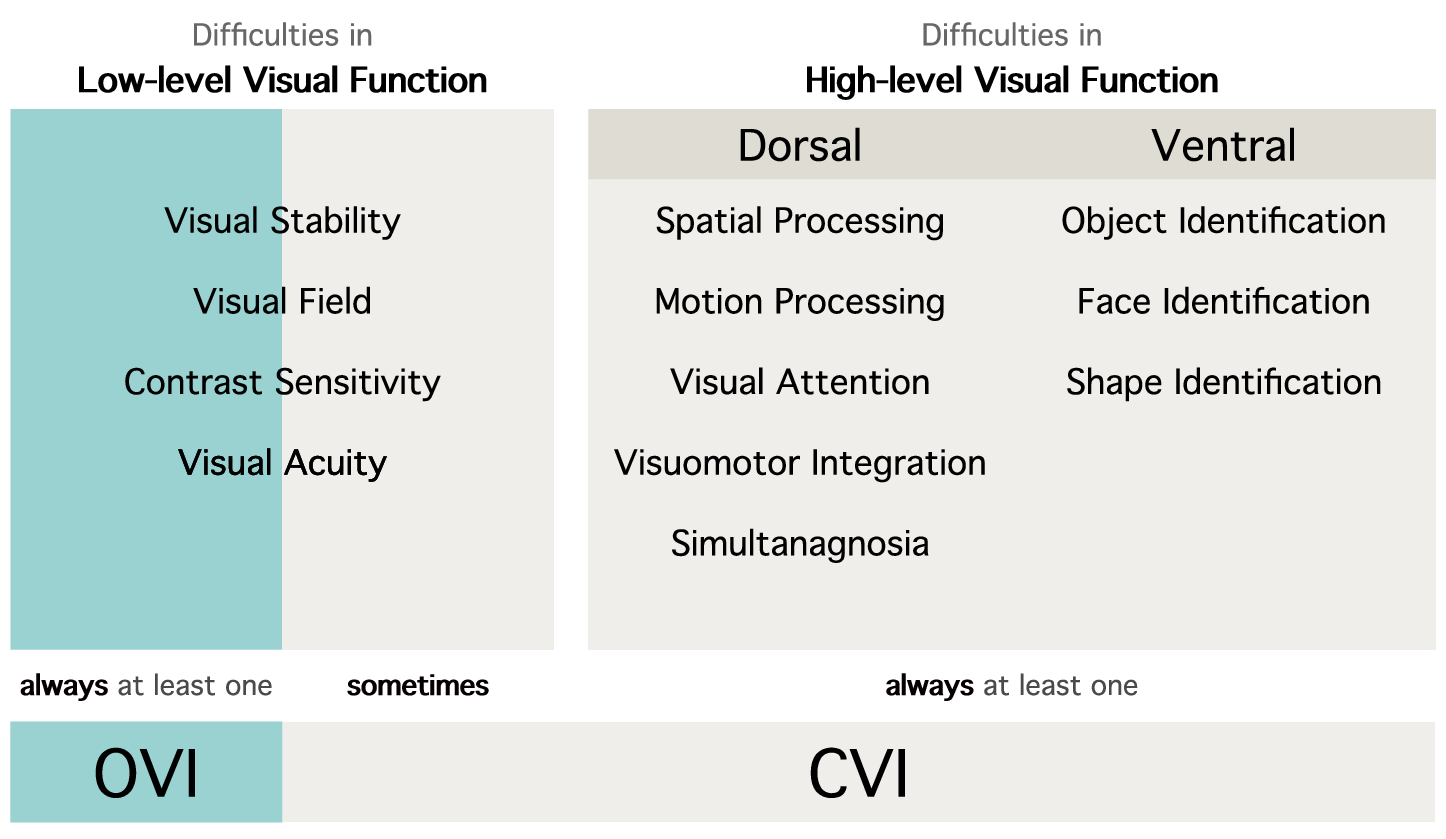}
    \caption{Overview of low-level and high-level visual difficulties for people with CVI. Note that this list is not exhaustive.}
    \Description{Two squares labelled “Difficulties in Low-level Visual Function’ and “Difficulties in High-level Visual Function”. Under the squares there is a horizontal bar indicating that at least one of low-level visual difficulties are always present in people with OVI, and sometimes for people with CVI, while at least one of the high-level visual difficulties are always present for people with CVI. The low-level visual difficulties are as follows: Visual Stability, Visual Field, Contrast Sensitivity, Visual Acuity. The high level visual difficulties are subdivided into two sub-categories - Dorsal and Ventral. The dorsal difficulties are as follows: Spatial processing, Motion processing, Visual Attention, Visuomotor Integration, Simultangnosia. The ventral difficulties are as follows: Object Identification, Face Identification, Shape Identification.}
    \label{fig:cvi-overview}
\end{figure*}

\section{Introduction}

Vision is not solely a product of the eyes; it also relies on intricate processes of the brain.
\textbf{Cerebral Visual Impairment (CVI)} is a visual dysfunction that is distinct from Ocular Vision Impairment (OVI), and caused by injury or disruption to the brain's visual processing centres \cite{sakki2018there}.
CVI has emerged as the predominant cause of childhood vision impairment in developed nations \cite{sandfeld2007visual, hatton2007babies, matsuba2006long}, affecting around 30-40\% of visually impaired children \cite{americanpediatric}.
As these children transition into adulthood, CVI is poised to become the leading cause of vision impairment \cite{bosch2014low}.

Recent breakthroughs in deep learning, particularly in computer vision and multimodal large language models~\cite{yin2023survey, vaswani2017attention, wu2023multimodal, chatgpt}, have sparked a surge of interest in  enhancing assistive technologies for people with vision impairments. 
Concurrently, a renewed interest in augmented and virtual reality has lead to the development of improved passthrough capabilities on most modern devices \cite{ar_trend, ricci2023virtual, munoz2022augmented, wu2021towards, masnadi2020vriassist}.
We define \textbf{Vision-Based Assistive Technologies (VBAT)} as devices incorporating machine learning, computer vision, image enhancement, and/or augmented/virtual reality to enhance the quality of life for people with vision impairments.
These can perform object detection, image enhancement and scene simplification and are built into smart glasses, smart phones or other wearables.
This study aims to address the primary question: \textbf{What is the current state of research at the intersection of CVI and VBAT, and what opportunities exist in this domain?}
The main contributions of our study are:
\begin{itemize}
    \item \textbf{Scoping Review:} We conduct a scoping review \cite{munn2018systematic} that provides insight into the current research landscape concerning CVI and technologies for VBAT. It reveals a paucity of research and a predominant focus on understanding CVI rather than assistance. 
    \item \textbf{Opportunities of VBAT:} We identify potential applications of VBAT that will significantly enhance the quality of life for people with CVI and propose promising areas for future work.
    \item \textbf{Raising Awareness of CVI in HCI and Assistive Technology (AT) Communities:} We discuss the importance and timing for the HCI and AT communities to address these opportunities.
\end{itemize}

We hope that our paper will provide guidance and inspiration to researchers in the fields of HCI and AT to address the unmet needs of the CVI community.

\begin{table*}[!h]
\centering
\caption{Similarities and Differences among people who are blind, have low vision, and have CVI}
\begin{tabular}{|p{3.6cm}|p{2cm}|p{4.7cm}|p{5cm}|}
\hline
\textbf{Criteria} & \textbf{Blindness} & \textbf{Low Vision}   & \textbf{CVI}          \\ \hline
\textbf{Interaction}       & Audio, Haptic               & Audio, Haptic, Visual (residual vision) & Audio, Haptic, Visual (normal vision) \\ \hline
\textbf{Visual Dysfunction}   & Low-Level                        & Low-Level             & Low-Level \& High-Level            \\ \hline
\textbf{Complexity Impact on \newline Vision}   & None                        & No Impact             & High Impact            \\ \hline
\textbf{Effects of Vision \newline Rehabilitation}   & Little Impact  & Little Impact     & High Impact            \\ \hline
\textbf{Association with Other Neurological Conditions}   & Rarely                        & Rarely             & Frequently             \\ \hline
\textbf{Visual Difficulties \newline Examples}   & Minimal to \newline no visual \newline perception                        &  Difficulty seeing objects, sensitive to contrast, identifying colour  & Finding objects in clutter, crowding of text, route finding, finding a \newline relative or friend in a crowded room            \\ \hline
\end{tabular}
\label{tab:differences}
\end{table*}

\section{Cerebral Visual Impairment (CVI)}
\label{sec: cvi}
Cerebral Visual Impairment (CVI) is the result of damage to the brain’s visual processing centres, rather than any physical damage to the eye \cite{lueck2015vision, roman2007cortical}. 
CVI is primarily observed in individuals with neurological conditions like cerebral palsy, stroke, or traumatic brain injury.
People with CVI face issues with interpreting and processing visual information, including difficulties in visual recognition, perception, understanding of their visual environment, and maintaining visual attention.

The terminology and definitions related to this condition, including Cerebral Visual Impairment \cite{philip2014identifying, lueck2015vision}, Cortical Vision Impairment \cite{whiting1985permanent, good1994cortical, roman2007cortical}, and Neurological Vision Impairment \cite{trobe2001neurology}, vary subtly and are a subject of ongoing debate 
~\cite{martin2016cerebral, mcdowell2020cvi}. 
We will use Cerebral Visual Impairment (CVI) to encompass a range of visual processing challenges resulting from damage to the brain's visual processing centres.

Ocular Visual Impairment (OVI) refers to vision impairments arising from problems within the eye or its associated structures, such as the retina, optic nerve, or cornea \cite{martin2016cerebral}. 
OVI examples include conditions like cataracts, glaucoma, and macular degeneration, which can cause a range of vision impairments from partial (low-vision) to total blindness.

\subsection{CVI Diagnosis and Prevalence}
\label{sec:morecvichildren}

In the last decade, our understanding of CVI has significantly expanded \cite{lueck2015vision, roman2007cortical}. 
Consequently, there has been an upsurge in the diagnosis of CVI among children.
This increased recognition can be attributed to the dissemination of knowledge regarding its primary causes, especially premature birth, among other factors \cite{bennett2020neuroplasticity, dutton2004association, good2001recent, kozeis2010brain, taylor2009differential}. 
CVI is frequently under-diagnosed, primarily because other forms of brain damage often coexist, causing a range of physical and cognitive impairments. 
In some instances, CVI might be mistaken for conditions like autism, learning disabilities, or behavioural challenges, further complicating accurate diagnosis \cite{swift2008cortical, williams2021cerebral}.

Numerous studies have focused on identifying children with CVI. 
Williams et al. \cite{williams2021cerebral} screened a substantial cohort of children and found a CVI prevalence rate of 3.4\% among those in mainstream classrooms. 
In addition to the large prevalence of CVI in children, it is believed that many adults grapple with undiagnosed CVI. 
Furthermore, adults can also develop CVI later in life, often stemming from factors such as strokes, traumatic brain injuries, 
and other neurological conditions \cite{lehmann2011basic, zhang2006homonymous}. 

At present, assessments and classifications of visual capability focus on OVI and primarily rely on criteria related to visual acuity and visual field. 
People with CVI may have normal visual acuity but struggle with higher-level visual functions~\cite{hyvarinen1995considerations, saidkasimova2007cognitive}.
For instance, a child with CVI may have difficulty recognising a parent only when in a crowded room, or finding a particular toy only when it's mixed with other objects \cite{dutton2003cognitive}. 
This presents clinicians with challenges when diagnosing people with CVI because the existing framework does not sufficiently address their needs.
It also means that many individuals with CVI do not meet the legal definition of blindness as per current standards \cite{kran2019cerebral}. 
This underscores the necessity for a broader understanding of CVI along with advocacy for a revision of the definitions of vision impairment to encompass people with CVI. 

\subsection{Low-Level Functional Vision vs. High-Level Functional Vision}
\label{sec:low-high-function}

People with CVI commonly experience difficulties in both low-level and high-level functional vision \cite{bennett2019assessment, chandna2021higher}.
Low-level functional vision encompasses fundamental aspects of visual perception like visual acuity, contrast sensitivity, visual field, and visual stability, similar to people with OVI. 

Conversely, high-level functional vision pertains to how individuals interpret and respond to visual information, including processes like visual recognition, comprehension, visual attention, and interaction with the environment. 
Studies indicate that these high-level vision impairments are associated with damage in two critical visual processing pathways in the brain--the Dorsal Stream and Ventral Stream \cite{bennett2020neuroplasticity}.

The two-stream model proposed by Goodale \cite{goodale2013separate}, consisting of the `dorsal stream'--connecting the occipital to the parietal cortex and the `ventral stream'--connecting the occipital to the inferior temporal cortex, offers a valuable framework for understanding higher-order visual processing difficulties in CVI \cite{haxby1991dissociation, mishkin1983object}.
Dysfunction in the dorsal stream typically manifests as vision impairments related to spatial and motion processing, difficulties in visual attention, challenges in visuomotor integration, and simultanagnosia (the inability to perceive more than one object at a time) \cite{bennett2020neuroplasticity, frcophth2009dorsal}. 
Meanwhile, damage along the ventral visual processing stream is associated with difficulties in object identification~\cite{bennett2020neuroplasticity, goodale2013separate, haxby1991dissociation}, such as recognising faces and shapes~\cite{andersson2006vision, houliston1999evidence}. 
However, dorsal and ventral stream difficulties often co-occur, suggesting that damage cannot be localised to a single brain area \cite{bennett2018assessing, dutton2011structured, macintyre2010dorsal}.
Figure \ref{fig:cvi-overview} provides an overview of the types of visual difficulties for people with CVI.

Recent advancements have provided tools to identify high-level visual function difficulties in people with CVI \cite{chandna2021higher, mcdowell2022using}.
These include questionnaires for initial assessment \cite{chandna2021higher}, and an iPad app that induces visual crowding \cite{mcdowell2020cvi, mcdowell2023validation}.

\subsection{Assistive Technologies for CVI}
\label{sec:assistivetechforcvi}

A large body of research within the field of HCI and AT has been dedicated to aiding people with OVI \cite{li2022scoping, patel2020assistive, simoes2020review, kuriakose2020multimodal, khan2021insight, valipoor2022recent, bashiri2018object, arora2019real, joshi2020yolo, chen2020smart, otaegui2013argus}.
These studies can be categorised into two primary areas: studies focusing on audio or tactile solutions for people who are blind (Ocular Blindness) \cite{abdolrahmani2021towards, kaul2021mobile, bhatia2022audio, villamarin2021haptic, yasmin2020haptic, srija2020raspberry}, and studies focusing on visual devices for people with low vision (Ocular Low Vision) \cite{zhao2015foresee, lang2020augmented, zhao2016cuesee, zhao2019designing, zhao2020effectiveness}.

Given the shared reliance on audio and haptic interactions in OVI and CVI, it is tempting to assume that research designed for OVI could be readily repurposed to benefit those with CVI.
This is especially true for Ocular Low Vision studies \cite{zhao2015foresee, lang2020augmented, zhao2016cuesee}, as they also encompass the visual modality as well.
However, this does not account for the two levels of functional vision difficulties within the CVI population.
In the case of people with low-level functional vision difficulties, we concur that devices and applications centered on OVI can indeed be applicable.
However, for people with high-level functional vision difficulties, evidence from the educational field indicates that strategies for people with OVI may not be effective \cite{martin2016cerebral}.
To elaborate, we have identified four main differences:
\begin{itemize}
    \item \textbf{Vision-Centric Interaction:} People with CVI 
    primarily rely on their vision, whereas people with low vision are more likely to prioritise audio and haptic interactions~\cite{zhao2020effectiveness}. 
    When combined simultaneously, multiple modalities, such as visual, audio, and haptic, can be overwhelming for people with CVI~\cite{philip2014identifying, lam2010cerebral}.
    \item \textbf{Complexity Impact on Vision:} 
    People with CVI experience escalating difficulties as visual complexity increases, while people with OVI maintain relatively stable performance \cite{bennett2018assessing}. Hence, the nature of device interaction becomes crucial, particularly in high-complexity tasks, as people with CVI require less taxing and overwhelming interactions compared to neurotypical and OVI participants.
    \item \textbf{Neuroplasticity and Brain Modification:} Unlike OVI, CVI is a brain-based disorder, enabling assistive devices to induce permanent changes in the brain's visual pathways through neuroplasticity~\cite{bennett2020neuroplasticity}. This effect can be both advantageous, potentially improving vision over time, or detrimental, possibly leading to lasting vision changes due to reliance on assistive devices. Therefore, researchers studying assistive devices for CVI should exercise caution and consider long-term effects through extended studies.
    \item \textbf{Frequent Association with Other Neurological Conditions:} In contrast to OVI, 
    CVI often co-occurs with other neurological conditions like cerebral palsy \cite{lueck2015vision}, which need to be considered within the research approach. 
\end{itemize}

Table \ref{tab:differences} provides a concise summary of the key similarities and distinctions among people who have ocular blindness, ocular low vision and CVI.

\section{Opportunities}
\label{sec: opportunities}

Recent advancements in artificial intelligence, particularly in computer vision and multimodal large language models~\cite{vaswani2017attention, yin2023survey, wu2023multimodal, chatgpt}, have sparked interest in developing assistive technologies for people with OVI \cite{li2022scoping, lee2021deep, assets2023}.
These range from basic object recognition \cite{bashiri2018object, arora2019real, joshi2020yolo}, to advanced autonomous guidance systems~\cite{chen2020smart, otaegui2013argus} and are built into devices from smart-glasses~\cite{guarese2021cooking, islam2020design, lin2020smart} to robotic dogs~\cite{hong2022development, bruno2019development}.
Furthermore, recent technological advancements in augmented and virtual reality \cite{ar_trend}, particularly in the passthrough capabilities of these devices, enable users to effortlessly overlay virtual objects and enhancements onto the physical world. 
This is also evident in the latest consumer devices like the Apple Vision Pro \cite{apple_vision_pro}, Meta Quest Pro \cite{meta_quest_pro}, and Meta Quest 3 \cite{meta_quest_3} marketing their passthrough capabilities.
This technology has already found diverse applications, including visual guidance \cite{vision_guidance}, compensation for color blindness \cite{vision_color_blind}, and visual noise cancellation through vision augmentation \cite{visual_noise}.
Studies have also investigated the use cases and effectiveness of augmented reality/virtual reality for people with OVI \cite{ricci2023virtual, munoz2022augmented, wu2021towards, masnadi2020vriassist}.

Considering the significance of vision-centric interaction for people with CVI, the question arises as to whether these technologies can be tailored to address their specific needs and work as assistive technologies.
Could these devices leverage a combination of machine learning, image enhancement, and augmented reality to provide real-time assistance to people with CVI? 
For example, VBAT could be used to fade out unwanted details when locating a pair of scissors, finding friends in a crowded venue, or finding a shop on a busy street.

Hence, given the opportunities for developing VBAT specifically tailored for the needs of people with CVI, in Section \ref{section: review} we dive into understanding the current research landscape in this domain.

\section{Literature Review}
\label{section: review}

Before delving further into the targeted research area (CVI and VBAT), we conducted an initial keyword search to gain insights into the research landscape pertaining to CVI. 
The primary objective was to obtain an overview of the prevailing trends in CVI studies. 
To achieve this, we executed a keyword search on Scopus using only the terms outlined in the first column ("Cerebral Visual Impairment") of Table~\ref{tab: keywords}. 
This search was conducted on January 2, 2024, resulting in a total of 2,464 papers.

We observed an upward trajectory in the number of papers, aligning with observations made by other CVI researchers \cite{lueck2015vision, roman2007cortical}. 
Approximately 85\% of the identified papers originated from disciplines closely associated with medicine, with over half exclusively within the medical domain. 
In stark contrast, a mere 4.9\% of the papers originated from the realms of computer science or engineering. 
This highlights the need for the HCI and AT communities to extend research on CVI beyond the medical approach.

\begin{table*}[t]
  \centering
 \caption{Scoping review keywords and criteria}
  \begin{subtable}{0.42\linewidth}
    \centering
            \caption{Keywords}
            \begin{tabular}{|p{4.15cm}|p{2.8cm}|}
            \hline
                \textbf{Cerebral Visual Impairment} & \textbf{Technologies for \newline VBAT} \\ \hline
                "cerebral visual impairment" & "computer vision" \\ 
                "cortical visual impairment" & "artificial intelligence" \\ 
                "neurological vision impairment" & "machine learning" \\ 
                ~ & "image enhancement" \\ 
                ~ & "augmented reality" \\ 
                ~ & "virtual reality" \\ 
                ~ & "mixed reality" \\ \hline
            \end{tabular}
            \label{tab: keywords}
  \end{subtable}
  \hspace{6mm}
  \begin{subtable}{0.5\linewidth}
    \centering
        \caption{Inclusion and Exclusion Criteria}
        \begin{tabular}{|p{8.5cm}|}
        \hline
        \textbf{Inclusion}                                                                  \\ \hline
        Papers focusing on CVI and technologies in Table \ref{tab: keywords}, irrespective of whether CVI participants are involved. \\
        Papers with CVI participants, regardless if focus is on CVI.   \\
        \hline
        \textbf{Exclusion}                                                                  \\ \hline
        Theses or Non-English papers. \\                
        Review papers and citations linked by references. \\
        Papers without any technologies in Table \ref{tab: keywords}. \\
        \hline
        \end{tabular}
        \label{tab:criteria}
  \end{subtable}
  \label{main-table}
\end{table*}

\subsection{Scoping Review}
\label{section: review2}

Given this information and the potential opportunities highlighted for Vision-Based Assistive Technologies (VBAT) for people with CVI in Section \ref{sec: opportunities}, our subsequent aim was to gain a comprehensive understanding of the research landscape within this specific domain.
To achieve this objective, we conducted a scoping review, examining the intersection of CVI with technologies for VBAT such as artificial intelligence, computer vision, image enhancement, and augmented/virtual reality. 
A scoping review was chosen as a method that is well-suited for examining research practices in a specific topic, identifying knowledge gaps, and categorising available evidence within a given field~\cite{munn2018systematic}.
After identifying these studies, our overarching goal was to sift through the literature and pinpoint a more focused subset of papers that specifically delved into the assistive technologies situated at the intersection of CVI and these technologies.

Our search was structured around two key criteria: Cerebral Visual Impairment and Technologies for VBAT.
The keywords, as displayed in Table~\ref{tab: keywords}, were combined using "OR" within the same column and "AND" across columns during the search process. 
We conducted this search across four databases: Google Scholar, Scopus, Web of Science, and PubMed. 
The search captured papers published up to January 2, 2024.

Our search yielded 595 papers from Google Scholar, 210 from Scopus, 9 from Web Of Science, and 5 from PubMed.
After consolidating and removing duplicates, a title and abstract screening process was conducted by one researcher with consultation from other members of the research team.
This led to 67 papers for full-text review.
Two researchers then reviewed the full texts using the criteria in Table~\ref{tab:criteria}, discussing and resolving any conflicts to identify a final set of 14 papers at the intersection of CVI and technologies for VBAT.

\subsubsection*{\textbf{Data Extraction:}}

The following information was recorded for each paper:
\begin{itemize}
    \item \textbf{Type of Study:} The primary focus of the study: Diagnosis (aiming to diagnose CVI), Understanding CVI (focused on understanding CVI), Simulation (simulating CVI conditions), Assistance (aimed at assisting users during use), or Rehabilitation (providing rehabilitation or therapy). These categories were determined by the two researchers conducting the full text review.
    \item \textbf{Involvement of CVI Participants:} Whether people with CVI were part of the study and, if so, whether their participation was during the design phase, evaluation phase, or both.
    \item \textbf{Type of Technology:} Specific type of technologies employed in the study.
    \item \textbf{Age Group and Other Details:} The participant age group: adults (over 18 years), children (below 18 years), both or unclear, and additional information, including publication venue, publication year, and authors' names.
\end{itemize}

\subsection{\textbf{Results}}

This section offers a summary of the main findings while the full dataset for all 14 papers is available in Table \ref{tab:full-paper-list} (in the Appendix).
Figure \ref{fig: cvi-papers-year} shows an increase in the publication trend over time, yet making definitive statements is challenging due to the limited number of papers.

\begin{figure*}[t]
\centering
\begin{subfigure}{.32\linewidth}
    \centering
    \includegraphics[width=4.5cm,keepaspectratio]{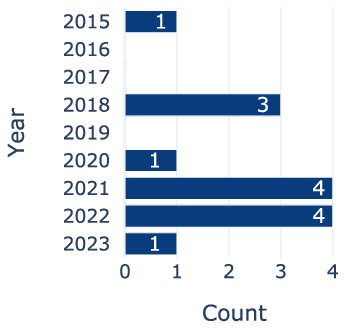}
    \caption{Publication Year}
    \Description{Horizontal bar chart with the count of papers for when the papers were published. The counts are as follows:2015 1, 2016 0, 2017 0, 2018 3, 2019 0, 2020 1 , 2021 4, 2022 4, 2023 1.}
    \label{fig: cvi-papers-year}
\end{subfigure}
\hfill
\begin{subfigure}{.32\linewidth}
    \centering
    \includegraphics[width=5cm,keepaspectratio]{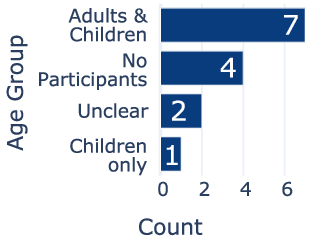}
    \caption{Age Group}
    \Description{Horizontal bar chart with the count of papers for the age groups of people involved in the studies. The counts are as follows: Adults \& Children 7, No participants 4, Unclear 2, Children only 1.}
    \label{fig:age-group}
\end{subfigure}
\hfill
\begin{subfigure}{.32\linewidth}
    \centering
    \includegraphics[width=5cm,keepaspectratio]{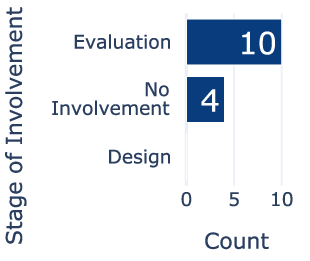}
    \caption{Stage of Involvement}
    \Description{Horizontal bar chart with the count of papers for stage of involvement. The counts are as follows: Evaluation 10, No Involvement 4, Design 0.}
    \label{fig:involvement-stage}
\end{subfigure}
\par\bigskip 
\par\bigskip 
\par\bigskip 
\begin{subfigure}{.44\linewidth}
    \includegraphics[width=8cm,keepaspectratio]{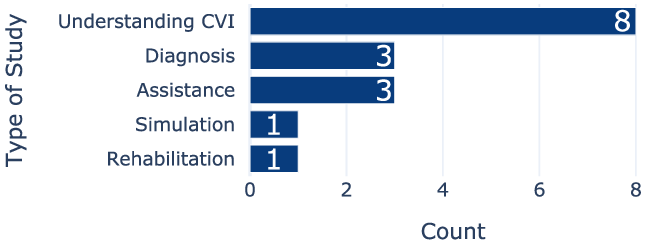}
    \caption{Type of Study}
    \label{fig:typeofstudy}
    \Description{Horizontal bar chart with the count of papers for each type of study. The types are sorted from highest to lowest count and are as follows: Understanding CVI 8, Diagnosis 3, Assistance 3, Simulation 1, Rehabilitation 1.}
\end{subfigure}
\hfill
\begin{subfigure}{.55\linewidth}
        \includegraphics[width=9.5cm,keepaspectratio]{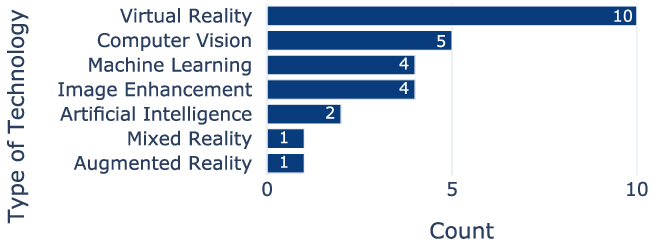}
        \caption{Type of Technology}
        \label{fig:technologies}
        \Description{Horizontal bar chart with the count of papers for each type of technology. These are sorted from highest to lowest count and are as follows: Virtual Reality 8, Computer Vision 5, Machine Learning 4, Image Enhancement 4, Artificial Intelligence 2, Mixed Reality 1, Augmented Reality 1.}
\end{subfigure}

\caption{Summary of findings from the Scoping Review}
\label{fig:review-2-summary}
\end{figure*}

\subsubsection*{Type of Study:}
Figure \ref{fig:typeofstudy} displays the distribution of study types extracted from the 14 papers. 
Some papers fell into multiple categories, hence the total count exceeds 14. 

We found only three VBAT papers, with just one directly addressing CVI requirements.
Birnbaum et al. \cite{birnbaum2015enhancing} suggested that presenting high contrast and low spatial frequency visual stimuli could increase visual awareness for people with CVI. 
They also suggested that augmented/virtual reality headsets could modify real-time visual input to enhance visual detection. 
However, their proposal was not implemented.

The other two assistance studies were not specifically focused on addressing concerns related to people with CVI \cite{pitt2023strategies, lorenzini2021personalized}.
Pitt and McCarthy \cite{pitt2023strategies} identified strategies for highlighting items within visual scene displays to support augmentative and alternative communication access.
They identified four methods for highlighting items: contrast based on light and dark, contrast based on colour, outline highlighting, and the utilisation of scale and motion.
In their future directions, they suggested the potential application of these strategies for people with CVI.
Lorenzini et al. \cite{lorenzini2021personalized} conducted a study with the objective to gather insights to reduce the likelihood of device abandonment when using portable head-mounted displays for telerehabilitation.
While the primary focus of their study was ocular low vision, participants with CVI were included in the study. 
Despite this inclusion, the study does not provide specific findings or discussions related to CVI.

Of the other studies, three focused on diagnosing CVI: Soni and Waoo \cite{soni2023convolutional} demonstrated the effectiveness of a convolutional neural network model for identifying and gaining insights into CVI.
Bennett and colleagues \cite{manley2022assessing, bambery2022virtual} developed two novel virtual reality based visual search tasks to objectively assess higher order processing abilities in CVI. 
Using the same tasks, they conducted many studies to expand the understanding of CVI.
All 8 studies identified for understanding CVI in the review were from Bennett and colleagues \cite{bennett2018virtual, bennett2018assessing, bennett2020alpha, pamir2021visual, bennett2021visual, federici2022altered, bambery2022virtual, zhang2022assessing}.

\subsubsection*{Involvement of CVI Participants:}

As depicted in Figure \ref{fig:age-group}, 10 out of the 14 papers incorporated CVI participants in the studies.
However, participant involvement was primarily in the studies that were focused on understanding CVI and diagnosis.
Such involvement in medical research is common practice, and in the context of these papers, participants were predominantly engaged for the purpose of validating hypotheses or effectiveness of diagnosis. 
None of the studies involved participants in the design or requirement gathering stages (as shown in Figure \ref{fig:involvement-stage}).

\subsubsection*{Type of Technology:}

Figure \ref{fig:technologies} offers an overview of the technologies employed in the examined papers. 
Virtual reality emerged as the predominant technology, largely due to the contributions by Bennett and colleagues. 
However, focusing on the subset of the three VBAT papers, computer vision, image enhancement, and machine learning were consistently proposed or employed. 
However, with only three papers, it is evident that these technologies are underutilised, presenting an opportunity for future studies to address this gap.

\subsubsection*{Age Group and Other Details:}
In the 7 studies including both adults and children (see Figure \ref{fig:age-group}), ages ranged from 7 to 28 years old. 
The focus on the pediatric and young adult population is understandable given the higher CVI prevalence in children (refer to Section \ref{sec:morecvichildren}). 
However, future studies should include participants across all age groups to better understand the diverse strategies employed by people living with CVI.

\section{Limitations, Future Work, and Conclusion}
It is challenging to draw definitive trends from only 14 papers, however this small number clearly demonstrates the significant lack of research in this important research area, with only three papers addressing VBAT. 

For future studies, we intend to adopt a human-centered approach, shifting away from a purely medical perspective. 
Our plan is to engage in a co-design processes with people living with CVI, exploring methods introduced by Pitt and McCarthy \cite{pitt2023strategies} along with other techniques discovered through the co-design process.

This paper underscored the unique requirements of people living with CVI, highlighted opportunities for VBAT, identified a significant research gap, and proposed two approaches to address it.
With recent technological advancements for VBAT and a wide-open field for research, now is the ideal time for researchers to create a substantial impact on people's lives.
As CVI emerges as a prominent vision impairment, we call upon researchers in the fields of HCI and AT to recognise and address this gap.


%

\begin{acks}
This project was funded by the Monash Data Futures Institute.
\end{acks}

\bibliographystyle{ACM-Reference-Format}
\bibliography{sample-base}

\appendix
\newpage
\section{REVIEW PAPERS}
\begin{table}[!h]
    \caption{List of 14 papers from the scoping review with the extracted data: Paper, Year, Age Group (A - Adults only, C - Children only, B - Both, ? - Unclear, N - None), Age Range ([Youngest] - [Eldest], U - Unclear, N - None), Type of Study (D - Diagnosis, U - Understanding of CVI, S - Simulation, A - Assistance, R - Rehabilitation), Involvement (D - Design Stage, E - Evaluation Stage, B - Both, N - None), Type of Technology} 
    \begin{tabular}{|p{0.05\textwidth}|p{0.04\textwidth}|p{0.02\textwidth}|p{0.02\textwidth}|p{0.02\textwidth}|p{0.02\textwidth}|p{0.165\textwidth}|}
        \hline
        \textbf{Study} & \textbf{Year} & \rotatebox[origin=c]{90}{\space \textbf{Age Group}} & \rotatebox[origin=c]{90}{\space \textbf{Age Range}} & \rotatebox[origin=c]{90}{\space \textbf{Type of Study}}  & 
        \rotatebox[origin=c]{90}{\space \textbf{Involvement} \space} & \textbf{Type of Technology} \\ \hline
        \cite{birnbaum2015enhancing} & 2015 & N & N & A & N & Computer Vision, \newline Image Enhancement, \newline Augmented Reality, \newline  Mixed Reality \\ \hline
        \cite{bennett2018virtual} & 2018 & N & 14-28 & U & E & Virtual Reality \\ \hline
        \cite{al20183d} & 2018 & N & N & S & N & Computer Vision, \newline  Artificial Intelligence, \newline  Machine Learning, \newline  Virtual Reality \\ \hline
        \cite{bennett2018assessing} & 2018 & B & 14-28 & U & E & Virtual Reality \\ \hline
        \cite{bennett2020alpha} & 2020 & B & 14-25 & U & E & Virtual Reality \\ \hline
        \cite{lorenzini2021personalized} & 2021 & ? & ? & A, R & E & Computer Vision, \newline Image Enhancement, \newline  Machine Learning \\ \hline
        \cite{pamir2021visual} & 2021  & C & 17 & U & E & Virtual Reality \\ \hline
        \cite{pitt2023strategies} & 2021 & N & N & A & N & Computer Vision, \newline Image Enhancement, \newline Machine Learning \\ \hline
        \cite{bennett2021visual} & 2021 & B & 15-22 & U & E & Virtual Reality \\ \hline
        \cite{federici2022altered} & 2022 &  B & 14-21 & U & E & Virtual Reality \\ \hline
        \cite{bambery2022virtual} & 2022 &  ? & ? & U, D & E & Virtual Reality \\ \hline
        \cite{zhang2022assessing} & 2022 &  B & 7-20 & U & E & Virtual Reality \\ \hline
        \cite{manley2022assessing} & 2022 &  B & 11-20 & D & E & Virtual Reality \\ \hline
        \cite{soni2023convolutional} & 2023 & N & N & D & N & Computer Vision, \newline  Artificial Intelligence, \newline  Image Enhancement, \newline  Machine Learning \\ \hline
        \end{tabular}
        \label{tab:full-paper-list} \\
\end{table}

\end{document}